\documentclass[twocolumn,showpacs,preprintnumbers,amsmath,amssymb]{revtex4}

\usepackage{graphicx}
\usepackage{dcolumn}
\usepackage{bm}

\begin{document}

\title{Adiabatic Condition for Nonlinear Systems}
\author{Han Pu$^{1}$, Peter Maenner$^{2}$, Weiping Zhang$^{3}$, and Hong Y.
Ling$^{2}$}
\affiliation{$^{1}$Department of Physics and Astronomy, and Rice Quantum Institute, Rice
University, Houston, TX 77251-1892, USA}
\affiliation{$^{2}$Department of Physics and Astronomy, Rowan University, Glassboro, New
Jersey, 08028-1700, USA}
\affiliation{$^{3}$Key Laboratory of Optical and Magnetic Resonance Spectroscopy
(Ministry of Education), Department of Physics, East China Normal
University,Shanghai 200062, P. R. China}

\begin{abstract}
We revisit the adiabatic criterion in stimulated Raman adiabatic
passage for the three-level $\Lambda$-system, and compare the
situation with and without nonlinearity. In linear systems, the
adiabatic condition is derived with the help of the instantaneous
eigenvalues and eigenstates of the Hamiltonian, a procedure that
breaks down in the presence of nonlinearity. Using an explicit
example relevant to photoassociation of atoms into diatomic
molecules, we demonstrate that the proper way to derive the
adiabatic condition for the nonlinear systems is through a
linearization procedure.
\end{abstract}

\date{\today }
\maketitle

According to the adiabatic theorem of quantum mechanics \cite{bohm51}, if
the Hamiltonian is changed sufficiently slowly, then a system in a given
non-degenerate eigenstate of the initial Hamiltonian (say, $|i\rangle $)
evolves into the corresponding eigenstate of the instantaneous Hamiltonian
without making any transitions. Here \textquotedblleft sufficiently
slowly\textquotedblright\ means that the rate of change of the Hamiltonian
is much smaller compared to the level spacings:
\begin{equation}
|\dot{H}|\ll \hbar |\omega _{fi}|^{2}\,,\;\;(\mathrm{for\,any}\,f\neq i)
\label{cond}
\end{equation}%
where $\omega _{fi}$ denotes the transition frequency between the
instantaneous eigenstates $|i(t)\rangle $ and $|f(t)\rangle $. Condition (\ref%
{cond}) is referred to as the adiabatic condition.

The standard version of the adiabatic theorem, which has many
applications in quantum state preparation and manipulation,
applies to linear quantum systems. We, however, have noticed that
the adiabatic condition (\ref{cond}) have been used in recent
studies of nonlinear quantum gases. This may be problematic since
the proof of the adiabatic condition makes explicit use of the
concept of an orthonormal set of energy eigenstates and the linear
superposition principle involving these states, both of which
become invalid when nonlinearity is introduced into the system.
The purpose of this paper is to provide a general method for
deriving the adiabatic condition in nonlinear systems. This is
achieved by considering a specific example of coherent population
transfer in three-level quantum
systems using the stimulated Raman adiabatic passage (STIRAP) method \cite%
{Gaubatz88,Kuklinski89}.
\begin{figure}
\centering
\includegraphics[width=3.in]{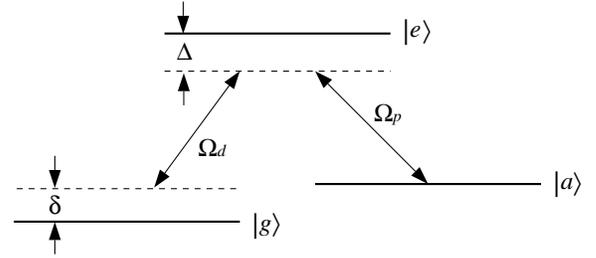}
\caption{The three-level system. $\Omega_p$ and $\Omega_d$ are the
two coupling strengths, $\Delta$ and $\delta$ are one and
two-photon detunings, respectively. } \label{lambda}
\end{figure}

Consider the three-level $\Lambda$-system schematically shown in Fig.~%
\ref{lambda}. The excited state $|e\rangle $ is coupled to two
ground states $|a\rangle $ and $|g\rangle $, with the coupling
strengths (Rabi frequencies) denoted as $\Omega _{p}$ and $\Omega
_{d}$, respectively. Here the subscripts \textquotedblleft
$p$\textquotedblright\ (\textquotedblleft $d$\textquotedblright )
stand for \textquotedblleft pump\textquotedblright\
(\textquotedblleft dump\textquotedblright ). In the interaction
picture, under the two-photon resonance condition ($\delta =0$),
the Hamiltonian of this system reads
\begin{equation}
H_{\mathrm{lin}}=-\hbar \Delta |e\rangle \langle e|+\frac{\hbar }{2}\left(
\Omega _{p}|a\rangle \langle e|+\Omega _{d}|g\rangle \langle e|+h.c.\right)
\,.  \label{hlin}
\end{equation}%
Without loss of generality, we will assume that the Rabi frequencies $\Omega
_{p,d}$ are real and positive.

Hamiltonian (\ref{hlin}) can be easily diagonalized. The
particular energy eigenstate
\begin{equation*}
|\mathrm{CPT}\rangle =\frac{1}{\Omega _{\mathrm{eff}}}\,\left( \Omega
_{d}\,|a\rangle -\Omega _{p}\,|g\rangle \right) \,,
\end{equation*}%
where $\Omega _{\mathrm{eff}}=\sqrt{\Omega _{p}^{2}+\Omega _{d}^{2}}$, is
known as the coherent population trapping (CPT) state or the dark state. The
state $|\mathrm{CPT}\rangle $ has zero eigenenergy and is decoupled from the
excited state. When $\Delta =0$, the other two eigenstates possess energies $%
\pm \hbar \,\Omega _{\mathrm{eff}}/2$. When $\Omega _{p,d}$ are varied
adiabatically, an initially prepared CPT state will remain in the
instantaneous CPT state. A straightforward application of (\ref{cond}) leads
to the adiabatic condition \cite{Kuklinski89}
\begin{equation}
r_{\mathrm{lin}}=\left\vert \frac{\dot{\Omega}_{p}\Omega _{d}-\dot{\Omega}%
_{d}\Omega _{p}}{\Omega _{\mathrm{eff}}^{3}}\right\vert =\frac{|\dot{\chi}|}{%
1+\chi ^{2}}\,\frac{1}{\Omega _{\mathrm{eff}}}\ll 1\,,  \label{con}
\end{equation}%
where $\chi \equiv \Omega _{p}/\Omega _{d}$. This property of the CPT state
facilitates a coherent population transfer from $|a\rangle $ to $|g\rangle $
when $\chi $ changes from 0 to $\infty $. This is the basis for STIRAP which
has wide applications ranging from chemical-reaction dynamics \cite{thermal}
to laser cooling of atoms \cite{kulin}.


In this paper, we want to derive the correct adiabatic condition
for a nonlinear three-level system describing a coupled
atom-molecule system. This system has recently received
significant attention due to the possibility of creating ultracold
molecules by associating cold atoms \cite{heinzen00,Mackie00}.

For the nonlinear system, we use the same level scheme of
Fig.~\ref{lambda}. Now state $|a\rangle $ represents an atomic
state, while $|e\rangle $ and $|g\rangle $ are excited and ground
diatomic molecular states. To keep things as simple as possible
without sacrificing essential physics, we neglect nonlinear
collisions between particles. The only nonlinearity comes from the
fact that it takes two atoms to form a molecule. Under the
two-photon resonance condition, the Hamiltonian in second
quantized form reads
\begin{equation}
\hat{H}_{\mathrm{nl}}=-\hbar \Delta \hat{\psi}_{e}^{\dag }\hat{\psi}_{e}+%
\frac{\hbar }{2}\left( \Omega _{p}\hat{\psi}_{e}^{\dag }\hat{\psi}_{a}\hat{%
\psi}_{a}+\Omega _{d}\hat{\psi}_{e}^{\dag }\hat{\psi}_{g}+h.c.\right) \,,
\label{hnl}
\end{equation}%
where $\hat{\psi}_{i}$ and $\hat{\psi}_{i}^{\dag }$ are the annihilation and
creation operators for state $|i\rangle $, respectively. In the mean-field
treatment, these are replaced by $c$-numbers $\psi _{i}$ and $\psi
_{i}^{\ast }$, which obey the following coupled equations:
\begin{subequations}
\label{dynamical equation}
\begin{align}
i\dot{\psi}_{a}& =\Omega _{p}\psi _{a}^{\ast }\psi _{e}\,, \\
i\dot{\psi}_{e}& =\Delta \psi _{e}+\frac{\Omega _{p}}{2}\psi _{a}^{2}+\frac{%
\Omega _{d}}{2}\psi _{g}\,,  \label{psim} \\
i\dot{\psi}_{g}& =\frac{\Omega _{d}^{\ast }}{2}\psi _{e}\,,
\end{align}%
\end{subequations}
$\psi _{i}$'s are normalized as $|\psi
_{a}|^{2}+2(|\psi _{g}|^{2}+|\psi _{e}|^{2})=1$, a consequence of
the conservation of total atom numbers.


To find the eigenstates of the system, we simply replace $id/dt$
at the left hand sides of Eqs.~(\ref{dynamical equation}) by
$\omega $, the eigenfrequency. It is easy to show, as in the
linear counterpart, the nonlinear $\Lambda $-system supports a CPT
eigenstate with zero eigenenergy \cite{Mackie00}, with the
corresponding state vector given by ${\boldsymbol \Psi }_{0}=(\psi
_{a}^{0},\,\psi _{e}^{0},\,\psi _{g}^{0})^{T}$ where
\begin{equation}
\label{CPT}
\psi _{a}^{0} =\left[ \frac{2\Omega _{d}}{\Omega _{d}+\Omega _{\mathrm{eff}%
}^{\mathrm{nl}}}\right] ^{1/2}\,, \; \psi _{e}^{0} =0\,, \;
\psi _{g}^{0} =-\frac{2\Omega _{p}}{\Omega _{d}+\Omega _{\mathrm{eff}}^{%
\mathrm{nl}}}\,.
\end{equation}
with $\Omega _{\mathrm{eff}}^{\mathrm{nl}}\equiv \sqrt{\Omega
_{d}^{2}+8\Omega _{p}^{2}}\,$. Evidence of the CPT state in
coupled atom-molecule systems has been reported in several recent
experiments \cite{Winkler05}.

Focusing on the case of vanishing single-photon detuning,
$\Delta=0$, we can easily obtain the other eigenstates:
\begin{equation*}
\frac{1}{\sqrt{2}}\,(0,\,\pm 1,\, 1)^T\,,
\end{equation*}
with eigenfrequencies $\pm \Omega_d/2$, respectively. Furthermore, when $%
\Omega_d/\Omega_p <1$, two more eigenstates appear:
\begin{equation*}
\left(\sqrt{\frac{1}{2} \left(1-\frac{\Omega_d^2}{\Omega_p^2} \right) }%
\,,\,\pm \frac{1}{2}\,,\, \frac{\Omega_d}{2\Omega_p} \right)^T\,,
\end{equation*}
with eigenfrequencies $\pm \Omega_p/2$, respectively. That there may exist
more eigenstates than the dimension of the Hilbert space is unique for
nonlinear systems \cite{liu05}. The CPT state distinguishes itself from
others with vanishing probability amplitude in the excited state $\psi_e$.
The non-orthorgonality between the CPT state and any of the other
eigenstates is quite obvious.

With the lack of a set of orthonormal energy eigenstates, the
conventional linear adiabatic condition (\ref{cond}) or
(\ref{con}) can no longer be used for the nonlinear system. We
want to derive the right adiabatic condition suitable for the
nonlinear $\Lambda$-system under study. Our goal is to find
the condition under which the system stays in the instantaneous CPT state (%
\ref{CPT}) when the Rabi frequencies $\Omega_p$ and $\Omega_d$ are
varied in time. To this end, we adopt the linear stability
analysis which has wide application in various nonlinear systems.

To begin with, we expand the state vector as:
\begin{equation*}
\psi _{i}=\psi _{i}^{0}+\delta \psi _{i}\,,
\end{equation*}%
where $\delta \psi _{i}$ represents the deviation of the probability
amplitude in state $|i\rangle $ from the CPT solution (\ref{CPT}).
Adiabaticity is obeyed as long as these deviations remain small. Inserting
this expansion into the dynamical Eqs.~(\ref{dynamical equation}), we obtain
the linearized equations:
\begin{equation}
i\frac{d}{dt}{\boldsymbol {\delta \psi} }={\bf M}\,{\boldsymbol
{\delta \psi }}+\dot{\boldsymbol \Psi }_{0} \,,\label{linear}
\end{equation}%
where \[ {\boldsymbol {\delta \psi} }=\left(
\begin{array}{c}
\delta \psi _{a} \\
\delta \psi _{e} \\
\delta \psi _{g}%
\end{array}%
\right)\,,\;\;\;{\bf M}=\left(
\begin{array}{ccc}
0 & \Omega _{p}\psi _{a}^{0} & 0 \\
\Omega _{p}\psi _{a}^{0} & \Delta  & \frac{\Omega _{d}}{2} \\
0 & \frac{\Omega _{d}}{2} & 0%
\end{array}%
\right)\,. \] The last term at the right hand side of
(\ref{linear}) represents the \textquotedblleft
source\textquotedblright\ term arising from
the temporal variation of Rabi frequencies ($%
\dot{\Omega}_{p}$ and $\dot{\Omega}_{d}$).

The eigenfrequencies of the linearized equations (i.e., the
eigenvalues of matrix ${\bf M}$) can be easily found as
\begin{eqnarray*}
\omega_0 &=& 0 \,, \\
\omega_\pm &=& \frac{1}{2} \,\left[ \Delta \pm \left( \Delta^2 +
\Omega_d\,\Omega_{\mathrm{eff}}^{\mathrm{nl}} \right)^{1/2} \right] \,.
\end{eqnarray*}
The corresponding eigenstates, i.e., the normal modes, are
\begin{equation*}
\mathbf{w}_0 = \mathcal{N}_0 \left(%
\begin{array}{c}
-\Omega_d/2 \\
0 \\
\Omega_p \psi_a^0%
\end{array}
\right)\,,\;\;\mathbf{w}_\pm = \mathcal{N}_\pm \left(%
\begin{array}{c}
\Omega_p \psi_a^0 \\
\omega_\pm \\
\Omega_d/2%
\end{array}
\right)\,,
\end{equation*}
where $\mathcal{N}_{0,\pm}$ are normalization constants. Note that $\mathbf{w%
}_{0,\pm}$ are orthogonal to each other.

Since the frequencies $\omega _{0.\pm }$ are all real, in the limit of $\dot{%
\Omega}_{p,d}\rightarrow 0$ (i.e., the vanishing source terms),
$\delta \psi _{i}$ will not grow in time and hence the system
initially prepared in a CPT state will stay in the instantaneous
CPT state. This is the adiabatic theorem generalized to this
nonlinear system. We remark that for a general nonlinear system,
complex eigenfrequencies of the linearized equations may appear
\cite{ling04}. These systems are \emph{dynamically unstable} such
that, even in the absence of source terms, the deviations may grow
from intrinsic quantum fluctuations \cite{liu05,wu00}.
Adiabaticity will therefore break down in the dynamical unstable
regimes of a nonlinear system.

The lack of complex eigenfrequencies in the current study guarantees that
the system under consideration is always dynamically stable, hence the
growth of deviations $\delta \psi_i$ can only be driven by the source terms.

From the first sight, the presence of the zero mode $\omega_0$
seems to be most worrisome, since the fluctuation in this mode is
energetically resonant
with the CPT state. Fortunately, as we now show, this resonance does \emph{%
not} lead to the growth of the zero mode. Writing the deviation vector as
superpositions of normal modes
\begin{equation*}
{\boldsymbol {\delta \psi} } = \sum_{\alpha=0,\pm}\,c_\alpha
\,\mathbf{w}_\alpha\,,
\end{equation*}
Equations (\ref{linear}) yield:
\begin{equation*}
i\dot{c}_\alpha = \omega_\alpha \,c_\alpha - i \,\mathbf{w}_\alpha^\dag \,
\dot{\boldsymbol \Psi}_0\,.
\end{equation*}
In particular, we have
\begin{equation*}
i \dot{c}_0 =- i \,\mathbf{w}_0^\dag \, \dot{\boldsymbol \Psi}_0 = -i
\mathcal{N}_0 \left( -\frac{\Omega_d}{2}\,\dot{\psi}_a^0 + \Omega_p
\psi_a^0\,\dot{\psi}_g^0 \right)\,,
\end{equation*}
Using the CPT solution given in Eqs.~(\ref{CPT}), it is
straightforward to show that the right hand side of the above
equation vanishes and hence the amplitude of the zero mode
fluctuation does not change in time. Therefore we reach a very
important conclusion: Although the zero mode fluctuation is
resonant with the CPT state, it is \emph{not} coupled by the
dynamics. From now on, we can simply ignore the zero mode and
focus on the remaining two modes $\omega_\pm$.

Assuming $c_\pm(0)=0$, the amplitudes $c_\pm (t)$ can be solved as
\begin{equation}
c_\pm(t)=-\int_0^t dt^{\prime}\, e^{i\omega_\pm (t^{\prime}-t)} \, \mathbf{w}%
_\pm^\dag(t^{\prime}) \dot{\boldsymbol \Psi}_0(t^{\prime}) \,.  \label{amp}
\end{equation}
Defining the adiabaticity parameter as
\begin{equation}
r_{\mathrm{nl}}(t) \equiv \frac{1}{2}\,\sqrt{|c_+(t)|^2 + |c_-(t)|^2}\,,
\label{r}
\end{equation}
which represents the population in the fluctuations above the CPT
solution, the adiabatic condition can now be defined
quantitatively as
\begin{equation}
r_{\mathrm{nl}}(t) \ll 1 \,.  \label{r<1}
\end{equation}

When $\omega_\pm$ is large, the exponential terms in (\ref{amp}) is rapidly
oscillating and the most significant contribution of the integral comes from
$t \approx t^{\prime}$. We may change the time variable of $\mathbf{w}%
_\pm^\dag(t^{\prime}) \dot{\boldsymbol \Psi}_0(t^{\prime})$ in the integrand
to $t$ and take it out of the integral, which yields
\begin{eqnarray*}
c_\pm(t) &=& - \mathbf{w}_\pm^\dag(t) \dot{\boldsymbol \Psi}_0(t)\,\frac{%
1-e^{-i\omega_\pm t}}{i\omega_\pm} \\
&=& - \mathcal{N}_\pm \, \frac{\dot{\Omega}_p \Omega_d - \Omega_p \dot{\Omega%
}_d}{\Omega_d + \Omega_{\mathrm{eff}}^{\mathrm{nl}}} \,\frac{%
1-e^{-i\omega_\pm t}}{i\omega_\pm}\,.
\end{eqnarray*}
The adiabatic condition now becomes
\begin{equation*}
r_{\mathrm{nl}}(t) \approx \frac{1}{2}\, \left(\frac{\mathcal{N}_+^2}{%
\omega_+^2} + \frac{\mathcal{N}_-^2}{\omega_-^2} \right)^{1/2}\,\frac{\left|%
\dot{\Omega}_p \Omega_d - \Omega_p \dot{\Omega}_d \right|}{\Omega_d +
\Omega_{\mathrm{eff}}^{\mathrm{nl}}} \ll 1 \,,
\end{equation*}
where we have neglected the oscillating term $e^{-i\omega_\pm t}$ in
evaluating $r_{\mathrm{nl}}(t)$.

For $\Delta =0 $, after some algebra, the above inequality leads to
\begin{equation}
r_{\mathrm{nl}}=\frac{|\dot{\chi}|}{1+ \sqrt{1+ 8 \chi^2}}\,\frac{1}{\Omega_{%
\mathrm{eff}}^{\mathrm{nl}}/2} \ll 1 \,,  \label{conn}
\end{equation}
where $\chi=\Omega_p/\Omega_d$ as before. Comparing (\ref{conn}) with (\ref%
{con}), we see that the adiabatic condition for the nonlinear
system has a somewhat similar expression to that for the linear
system, with, however, the following major difference: $\chi^2$ in
the denominator
at the left hand side of (\ref{con}) is replaced by $\sqrt{1+ 8 \chi^2}$ in (%
\ref{conn}). It follows that, at the later stage of STIRAP where $\chi \gg 1$%
, the adiabatic condition for the nonlinear system is harder to fulfil than
for its linear counterpart.

We remark that for the linear system, we can use the same linearization
procedure to derive the adiabatic condition. It is obvious that, in this
case, the eigenfrequencies of the linearized equations are the same as the
ones of the original equations. Eqs.~(\ref{r}) and (\ref{r<1}) will then
reproduce the linear adiabatic condition (\ref{con}).

To support the above analysis, we use the following numerical example. We
choose two equal-amplitude equal-width Gaussian pulses for the pump and dump
field centered at $t_{p,d}$, respectively,
\begin{equation*}
\Omega _{p,d}(t)=\Omega _{0}\,e^{-(t-t_{p,d})^{2}}\,,
\end{equation*}%
where we have taken the pulse width as the units for time. A
counter-intuitive pulse sequence \cite{Hioe83} requires that $t_{d}<t_{p}\,$%
. At the initial time ($t=0$), only the $|a\rangle $ state is populated,
i.e.,
\begin{equation*}
\psi _{a}(0)=1\,,\;\;\psi _{g}(0)=\psi _{m}(0)=0\,.
\end{equation*}%
We evolve the state vector under the Hamiltonian $H_{\mathrm{lin}}$ and $H_{%
\mathrm{ln}}$ for the linear and nonlinear systems, respectively.
\begin{figure}
\centering
\includegraphics[width=2.6in]{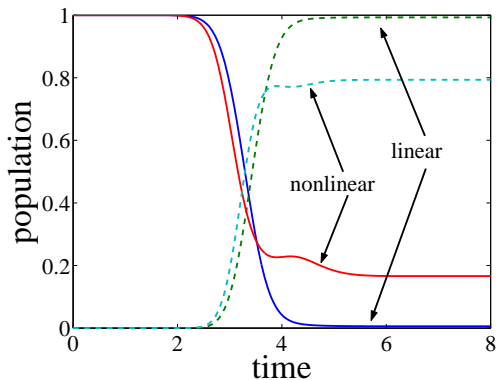}
\caption{(Color online) Population dynamics. Solid lines:
population in state $|a\rangle$ ($|\psi_a|^2$); dashed lines:
population in state $|g\rangle$ ($|\psi_g|^2$ for the linear case
and $2|\psi_g|^2$ for the nonlinear case). The parameters used
here are $\Delta=0$, $\Omega_0=5$, $t_d=3$ and $t_p=3.8$.}
\label{population}
\end{figure}

Figure \ref{population} shows an example of the population
dynamics. There are not much differences at the early stage for
the linear and nonlinear cases. However, at the later stage of the
STIRAP, the nonlinear system can no longer follow the CPT state
while the linear system stays closely in the CPT state. In the
end, we achieve an 80\% population transfer in the former and near
perfect transfer in the latter.
\begin{figure}
\centering
\includegraphics[width=2.6in]{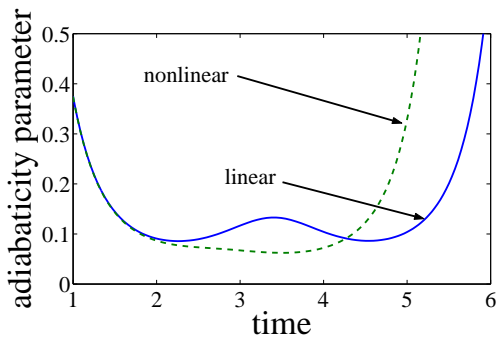}
\caption{(Color online) Adiabaticity parameter as defined in
(\ref{con}) and (\ref{conn}) for the linear and nonlinear systems,
respectively. Same parameters as in Fig.~\ref{population}. }
\label{adiabaticity}
\end{figure}

Figure \ref{adiabaticity} shows the corresponding time evolution of the
adiabaticity parameter. At the early stage of STIRAP, the two parameters in
the linear and nonlinear systems are almost identical. However, at the later
stage, we have
\begin{equation*}
r_{\mathrm{nl}}\gg r_{\mathrm{lin}}\,,
\end{equation*}%
in perfect agreement with our analysis, which also explains the reduced
population transfer efficiency for the nonlinear system and the population
dynamics as demonstrated in Fig.~\ref{population}.

In conclusion, we have shown that, for nonlinear systems, the
textbook method used in deriving the adiabatic condition fails to
provide the correct answer, and the proper way to derive the
adiabatic condition is through a linearization procedure. This is
demonstrated by considering the coherent population transfer process
via STIRAP in both a linear and a nonlinear three-level
$\Lambda$-system, which are indeed governed by different adiabatic
conditions. As a side, our analytical result of the nonlinear
adiabatic condition (\ref{conn}) will become very useful in
determining the required laser parameters for photoassociation via
STIRAP.

As we have mentioned, in order to keep the mathematics as simple
as possible, we have neglected here the nonlinear collisions
between particles which are important in the ultracold quantum
degenerate atomic/molecular systems. The inclusion of the
collisions will introduce dynamically unstable regimes for the CPT
states \cite{ling04}. With collisions, the linearization procedure
is equivalent to the Bogoliubov treatment of weakly interacting
Bose condensates. However, the orthonormality of linearized
eigenstates is replaced by \emph{bi-orthonormality} of the
Bogoliubov normal modes, and the zero-frequency mode corresponds
to the Goldstone modes related to the spontaneously broken
symmetries. These changes lead to considerably more complicated
algebra. The adiabatic condition for collisional systems will be
studied elsewhere.

This work is supported by the US National Science Foundation (HP
and HYL), by the National Natural Science Foundation of China
under Grant No. 10474055 and No 10588402, and by the Science and
Technology Commission of Shanghai Municipality under Grant No.
04DZ14009 and No. 05PJ14038 (WZ).

\bigskip

\end{document}